\documentstyle[12pt]{article}
\begin{document}

\begin{center}
{\bf  QUANTUM MECHANICS AS KINETICS}
\end{center}
\begin{center} {\bf L V Prokhorov}  \end{center}

\begin{center}
{V.A.Fock Institute of Physics \\
Sankt-Petersburg State University, Russia} \\
\end{center}

\begin{center}
{\bf  Abstract} \\
\end{center}
Relations between Hamiltonian mechanics and quantum mechanics are studied.
It is stressed that classical mechanics possesses all the specific features
of quantum theory: operators, complex variables, probabilities (in case of
ergodic systems). The Planck constant and the Fock space appear after
putting a dynamical system in a thermal bath. For harmonic oscillator in a
thermal bath, the probability amplitudes can be identified with the
complex valued phase functions $f(q+ip)$ describing small deviations from
the equilibrium state, when the time of relaxation is large. A chain of such
oscillators models both the one-dimensional space (or string), and
one-dimensional quantum field theory.

\noindent
\section{\bf Introduction}

At the present time it becomes clear that to uncover the nature of quantum
description one has to turn to the Planck scales $l_{\rm P}\sim 1.6\cdot
10^{-33}$ cm [1,2]. For example, 't Hooft came to conclusion that to
construct self-consistent quantum gravity one should 1) admit discrete
space-time and physical variables, 2) postulate at the Planck
scales deterministic classical mechanics (CM), 3) admit (locally)
dissipative processes [1]. Of course, at the atomic level one cannot
obtain quantum mechanics (QM) from the classical one. It may be possible
only in the models where matter, mechanics and space appear
simultaneously [2]. For the recent activity in this direction see
[3,4].  The problem of the Planck constant (what is the nature of $h$?) was
not discussed in [1,3,4].

The aim of the article is to show that evolution of non-equilibrium states
of a harmonic oscillator in a thermal bath is described by probability
amplitudes (if the deviation from the equilibrium is small, and the
relaxation time is large). It is the heat bath that is responsible for
appearance of probabilities and the Planck constant $h$. A chain of such
oscillators models one-dimensional quantum field theory or quantized bosonic
string. One-particle excitations of the field play the role of particles and
are described by wave functions. The distances between the oscillators
are supposed to be of order $l_{\rm P}$. Thus, $1D$ relativistic quantum
world can be built in the framework of non-relativistic classical mechanics.
The harmonic oscillator plays the prominent role in this model (as is
well known fields are ordered sets of harmonic oscillators).

In section 2 the attention is drawn to the fact that Hamiltonian mechanics
possesses all the specific properties of quantum theory: operators,
complex valued functions, probabilities (for ergodic systems). Moreover, in
case of complex systems there appears also a constant with dimension of
action and the Hilbert space.

In section 3 equilibrium and non-equilibrium distributions for a system in
a thermal bath are discussed. Variations of canonical variables are
divided into two classes. Those, preserving the Gibbs distribution satisfy
equations analogous to the Hamiltonian equations of motion. All the other
variations give rise to non-equilibrium states.

Study of harmonic oscillator in a thermal bath (section 4) shows that phase
functions $f(z)$, $z=(q+ip)/\sqrt 2$, describing non-equilibrium states, in
case of large relaxation time $t_r$ compose the Fock space and may be
identified with the probability amplitudes.

In section 5 a chain of harmonic oscillators in a heat bath is considered.
It is shown that in the continuum limit one obtains 1D quantum field
theory. It elucidates the nature of quantum fields, matter (excitations of
the fields) and physical space (the chain of oscillators).

In Conclusion (section 6) the main results of the paper are summarized, and
the consequence of the proposed model for the Universe is pointed out
(disappearance of coherent excitations of the vacuum, i.e. "disappearance of
matter").

\section{\bf Hamiltonian mechanics and mathematical apparatus of quantum
mechanics}

To define Hamiltonian mechanics, one has to define 1) phase space (PS) --- an
even-dimensional manifold, 2) non-degenerate closed symplectic form on PS,
3) the Hamilton function $H(q,p)$. The symplectic form

\begin{equation}
\omega^2(q,p)=\sum_{k=1}^n\omega^{-1}_k(q,p)dq_k\wedge dp_k
\end{equation}
defines the Poisson bracket for phase functions $f(q,p),g(q,p)$
\begin{equation}
\{f,g\}=\sum_{k=1}^n\omega_k(q,p)
\left(\frac{\partial f}{\partial q_k}\frac {\partial g}{\partial
p_k}-\frac{\partial f}{\partial p_k}\frac{\partial g}{\partial
q_k}\right)\equiv \sum_{k=1}^n\omega_k(q,p)\frac{\partial (f,g)}{\partial
(q_k,p_k)}.
\end{equation}
Because we are going to use non-canonical transformations, the following
symbol is proved to be useful
\begin{equation} \{f,g/q,p\}=\frac{\partial (f,g)}{\partial (q,p)}.
\end{equation}

The equation of motion reads
\begin{equation}
\dot f =\{f,H\}.
\end{equation}
Hamiltonian mechanics contains all the essential constituents of quantum
mechanics. For the readers convenience we give some details used in what
follows.

{\bf Operators.} Equation (4) can be rewritten in the form

\begin{equation}
\dot f = \{H,\; /p,q\}f\equiv\hat H_{cl}f,
\end{equation}
where $\{H,\; /p,q\}$ is an operator: $\{H,\;/p,q\}f= \{f,H/q,p\}$. For
simplicity, we took $n=1$ and $\omega(q,p)=1$. $\hat H_{cl}$ is an operator
in CM [5].

{\bf Complex functions.} Let us take as an example harmonic oscillator with
the Hamiltonian

\begin{equation}
H=\frac{1}{2}(\frac{\tilde p^2}{m}+\gamma \tilde q^2)
=\frac{\omega}{2}(p^2+q^2), \ \ \omega=\sqrt \frac{\gamma}{m},
\ \ p^2 = \frac{\tilde p^2}{\sqrt{\gamma m}}, \ \ q^2 = \tilde q^2 \sqrt
{\gamma m}.
\end{equation}
The equations of motion are
\begin{equation}
\dot {\bf x}(q,p)=\omega \hat {J} {\bf x}(q,p), \ \ \hat J =
\left( \begin{array}{cc} 0&1 \\ -1&0 \end{array} \right), \ \ \hat J^2=-1
\ \  \ \ ({\bf x}(q,p): \; x_1=q, x_2=p).
\end{equation}
In normal coordinates
$z=(q+ip)/\sqrt 2, \bar z=(q-ip)/\sqrt 2$ we have
$$ \dot {\bf x}(\bar z,z)=i\hat {S}{\bf x}(\bar z,z), \ \ {\bf x}(\bar
z,z)= \hat {U}{\bf x}(q,p), \ \ \hat {S}=\left( \begin{array}{cc} 1&0 \\
0&-1 \end{array} \right), \ \ \hat {U}=
\frac{1}{\sqrt 2}\left( \begin{array}{cc} 1&-i \\ 1&i \end{array} \right),
\ \ \hat U^+\hat U=1,$$
i.e. Eqs. (7) decouple, and one may use e.g. only the second equation
\begin{equation}
\dot z = -i\omega z;
\end{equation}
the first one can be obtained by complex conjugation. Complex canonical
variables $z,\bar z$ are useful in general case [6,7].

{\bf The imaginary unit.} Notice, that 1) matrix $\hat J$ in (7) is
transformed into the imaginary unit $i$ in (8): $\hat U\hat J\hat U^+=i\hat
{S}$, 2) transformation $q,p\rightarrow z,\bar z$ {\it is not canonical
one}: $\{\bar z,z/q,p\}=i$ (though $\hat U$ is a unitary matrix; for a
canonical transformation $q,p \rightarrow Q,P$ by definition
$\{Q,P/q,p\}=1$).

{\bf Probabilities.} Complex ergodic systems are stochastic ones; the time
average of phase functions is equal to their average over an ensemble. Thus,
in deterministic classical theories there appear probabilities and (for
subsystems, see e.g. [8]) the Gibbs distributions. This is one of the most
important points of our consideration.

{\bf Hilbert space.} The Gibbs distribution allows to define the Hilbert
space; for complex valued phase functions $f(q,p),g(q,p)$ one can introduce
the inner product [5,9]
\begin{equation}
(f,g)=Z^{-1}\int\bar f ge^{-\beta H(q,p)}d^n q d^n p
\end{equation}
(here $\beta =1/kT$, $k$ --- the Boltzmann
constant, $T$ --- the temperature, $Z$ --- the normalizing constant). It was
stressed however [9], that in QM wave functions are defined in configuration
(or momentum) space, not in PS. The connection between the Hilbert space
with the inner product (9) and the Fock space was not noticed. The point is:
arbitrary phase function $f(q,p)$ cannot be a wave function because
$\hat q$ and $\hat p$ do not commute. But the phase functions $\tilde
f(q,p)\equiv f(z)$ can be considered (and are considered) as wave functions
in the Fock space --- now canonical variables are $\bar z,z$
(see section 4).

{\bf The Planck constant $h$.} For finite motion the integral
\begin{equation}
Z=\int e^{-\beta H(q,p)}d^n q d^n p
\end{equation}
exists. Then, the constant
\begin{equation}
Z^{1/n}=h
\end{equation}
has the dimension of action, and in certain cases it can be identified with
the Planck constant $h$ (see sections 4,5). Of course, in general case this
$h$ depends on $H,\beta$ and is not a universal constant. But in a world
made of identical elements, e.g. in 1D quantum field theory (linear chain of
oscillators) this constant is universal one.

\section{\bf Equilibrium and non-equilibrium distributions}

{\bf The Gibbs distribution and the Hamiltonian equations of motion.} As was
already mentioned, classical mechanics of complex ergodic systems is
stochastic and bears the Gibbs distribution. Problem: find variations of
canonical variables preserving the Gibbs distribution. From $\delta
\exp(-\beta H)=0$ one gets

\begin{equation}
\delta H(q,p)=\sum_i \left(\frac{\partial H}{\partial q_i}\delta q_i
+\frac{\partial H}{\partial p_i}\delta p_i \right)\equiv \sum_i \nabla_i H
\delta {\bf x}_i \equiv \nabla H\delta {\bf x}=0
\end{equation}
(${\bf x}_i={\bf x}(q_i,p_i), \; {\bf x}={\bf x}(q_1,p_1,...,q_n,p_n)$).
Solution of this equation is

\begin{equation}
\delta {\bf x} =\hat {\Omega} \nabla H \delta t,
\end{equation}
where $\hat {\Omega}$ is some $2n\times 2n$ antisymmetric matrix, such that

\begin{equation}
\nabla H\hat {\Omega} \nabla H =0.
\end{equation}
Among solutions (13) there are solutions of the Hamiltonian
equations of motion.  In the simplest case ($\hat {\Omega}= \stackrel
{n}{\otimes}\hat J,\ \ \omega_k(q,p)=1$) equations (13) are nothing but the
familiar Hamiltonian equations

\begin{equation}
\dot q_i =\frac{\partial H}{\partial p_i},\ \ \dot p_i =-\frac{\partial
H}{\partial q_i}.
\end{equation}
It follows from (12) that an arbitrary variation $\delta {\bf x}$ is the sum

\begin{equation}
\delta {\bf x}=\delta {\bf x}_{\perp}+\delta {\bf x}_{\parallel},
\end{equation}
where $\delta {\bf x}_{\perp}$ are given by (13). Variations $\delta {\bf
x}_{\perp}$ preserve the Gibbs distribution while variations $\delta {\bf
x}_{\parallel}$ give rise to non-equilibrium states.

{\bf Non-equilibrium distributions.} The Gibbs distribution for harmonic
oscillator introduces measure in PS
\begin{equation}
d\mu (\bar z,z)=\frac{d\bar z\wedge dz}
{ih}e^{-\bar z z/\hbar},\ \ \hbar= \frac{h}{2\pi}=\frac{1}{\beta \omega}
\end{equation}
(see (6), (11)). Any other measure $\mu_p\neq\mu$ describes some
non-equilibrium state
\begin{equation}
d\mu_p=(d\mu_p/d\mu)d\mu\equiv p(\bar z,z)d\mu,\ \ p(\bar z,z)\geq 0.
\end{equation}
Now, consider variation of $\bar z,z$ with small non-zero $\delta {\bf
x}_{\parallel}$ in (16); we have
$$\omega\bar z z\rightarrow \omega\bar z z - \beta^{-1}(cz + \bar c\bar z) +
...,\quad e^{-\beta\omega\bar z z}\rightarrow |e^{c z + ...}|^2 e^{-\bar z
z/\hbar},$$
and
\begin{equation}
d\mu(\bar z,z)\rightarrow d\mu_f(\bar z,z)=|f(z)|^2d\mu(\bar z,z), \ \
f(z)= f_c(z)=e^{c z};
\end{equation}
the higher degrees of $\bar z,z$ in the exponential are omitted. Then,
instead of $p(\bar z,z)$ in (18) we have $|f(z)|^2$, so {\it to study
evolution of a non-equilibrium state it is enough to consider evolution of}
$f(z)$. Evolution of a single analytical function $f(z)$ describes evolution
of a non-equilibrium state; entire functions $f(z)$ are elements of the Fock
space [10,11]. This is the key to the problem.

{\it Remark.} Introduction of $h$ by Eqs. (11), (17) may look unconvincing
at the first sight. Notice, however, that (17) can be obtained by projection
of a sphere $(\varphi,\theta)$ on complex plane $z$
$$
|z|^2=\frac{1}{\beta}\ln\frac{1}{2\beta R^2
\sin^2\theta/2},\ \ \arg z= \varphi;\quad
R^2\sin \theta d\varphi\wedge d\theta = d\mu (\bar z,z),
$$
where $R$ is radius of the sphere [2]. If PS is a sphere then its area can
be identified with $h=4\pi R^2$ --- the universal constant with dimension of
action for Hamiltonian dynamics on the sphere. A chain of such oscillators
can model a bosonic string, while a network of strings models e.g. $3D$
space [2]. Hamiltonian mechanics in this space possesses the universal
constant $h$.

\section{\bf Harmonic oscillator in a thermal bath and quantum mechanics}

It is easy to show now that for harmonic oscillator evolution of small (in
the sense of section 3) deviations from the equilibrium state is described
by quantum mechanics if time of relaxation is large ($t_r\gg \omega^{-1}$).

{\bf Evolution of non-equilibrium states.} Hamiltonian (6) in complex
variables takes the form
\begin{equation}
H=\!\frac{\omega}{2} \left( \begin{array}{c} q \\ p \end{array}
\right) \left( \begin{array}{cc} 1&0 \\ 0&1 \end{array} \right) \left(
\begin{array}{c} q \\ p \end{array} \right)=\!
\frac{\omega}{2} \left( \begin{array}{c} \bar z \\ z \end{array}
\right)U^*U^+\left(\begin{array}{c} \bar z \\ z  \end{array} \right) =\!
\frac{\omega}{2} \left( \begin{array}{c} \bar z \\ z \end{array}
\right)\left( \begin{array}{cc} 0&1 \\ 1&0 \end{array} \right)
\left(\begin{array}{c} \bar z \\ z  \end{array} \right)\! =\!
\frac{\omega}{2}(\bar z z + z \bar z),
\end{equation}
and the equation of motion for $f(z)$ reads
\begin{equation}
\dot f=\{f,H/q,p\}=i\{f,H/\bar z,z\}=-i\omega
z\frac{d}{dz}f(z)=-i\omega \frac{d}{d\ln z}f(z)\equiv \hat H_{cl}f
\end{equation}
or, in terms of $\varphi =\arg z$
\begin{equation}
(\partial_t+\omega\partial_\varphi)f(\varphi,t)=0
\end{equation}
($d|z|/dt=0$). Functions $f(\varphi-\omega t)$ solve equation (22); they
describe waves on the circle $|z|=const$ and should be periodic functions.
So, $f(z)$ can be developed into the Laurent series. But in general case
such functions are singular at $z=0$ and do not suit for description of
any sensible state of harmonic oscillator. The proper functions are those
given by the Maclaurin series

\begin{equation} f(z)=\sum\limits_{n=0}^{\infty}c_n Z_n(z),\quad
Z_n(z)=z^n/\sqrt{n!}.
\end{equation}
Functions $Z_n$ satisfy equations of motion
\begin{equation} \dot Z_n=-i\omega nZ_n,
\end{equation}
and evidently correspond to motion with frequencies $\omega n$.

{\bf Probability amplitudes.} Solutions of equations (21), (22) form a linear
space. Now, define the scalar product
\begin{equation}
(g,f)=\int d\mu(\bar z,z)\overline{g(z)}f(z);
\end{equation}
then the entire functions $f(z)$ of order $\rho\leq 2$ form the Fock space
${\cal F}$ [10,11], i.e. the Hilbert space of harmonic oscillator. We
conclude: the phase functions $f(z)$, describing non-equilibrium
distributions of harmonic oscillator, can be identified with the probability
amplitudes. For example, it is well known [10,11] that functions $Z_n(z/\sqrt
\hbar)$ give the orthonormal basis in the Hilbert space

\begin{equation}
(Z_n,Z_m)=\delta_{nm},
\end{equation}
and we demand

\begin{equation}
(f,f)=\sum\limits_{n=0}^{\infty}|c_n|^2=1.
\end{equation}
The latter equality means that $|f(z)|^2$ may be identified with the
probability density. Numbers $|c_n|^2$  also represent some (discrete)
probability distribution. Evolution of a non-equilibrium distribution (i.e.
of $|f(z)|^2$) is given by evolution of a single function $f(z)$.

It is easy to see that smallness of $c$ in (19) does not impose limitations
on the set of functions $e^{cz}$ as a basis in ${\cal F}$. It is known [10]
that for arbitrary small $\epsilon >0$ vectors $e^{c_kz},\: |c_k|<\epsilon$
form a (over)complete basis in ${\cal F}$ if an infinite sequence of points
$c_k$ in the complex plane converges to some limit $c_0$ (one can omit
from the sequence any finite number of points). It means that if some
function $\psi(z)\in {\cal F}$ is orthogonal to all the functions $e^{c_kz}$
then $\psi(z)=0$. Indeed, functions $e^{\bar c z}=\langle z|\bar c\rangle=
\bar c(z)$ play the role of $\delta$-functions in ${\cal F}$, and
$(\bar c_k,\psi)=\langle c_k|\psi \rangle=\psi (c_k)=0, \forall k$. But
$\psi(z)$ is entire function, so $\psi(z)=0$.

{\bf Commutation relations.} It is well known that for the Fock space with
measure (17) operator $\hat{\bar z}$ defined as $\hat{\bar
z}\overline{g(z)}=\bar z\overline{g(z)}$ has the Hermitian conjugate $\hat
z=\hbar d/d\bar z$ ($\hat {z}f(z)=zf(z)$; it can be proved by integration by
parts in (25) [10]).  The commutation relation
\begin{equation} [\hat z,\hat{\bar z}]=\hbar
\end{equation}
allows to identify $\hat{\bar z},\hat z$ as the creation and annihilation
operators for quantized harmonic oscillator. Then, the canonical variables
$q,p$ also become operators
\begin{equation} [\hat q,\hat p]=\frac{i}{2}[\hat{\bar z}+\hat z,\hat{\bar
z}-\hat z]=i\hbar.
\end{equation}
Remember, that we are working in the framework of pure classical theory.

{\bf The Schroedinger equation.} The classical equation of motion (21) can
be identified with the Schroedinger equation if we multiply it by $i\hbar$:
\begin{equation}
i\hbar\dot Z_n(z)= \hat H_{Cl} Z_n(z),\quad \hat H_{Cl}=i\hbar \hat H_{cl}=
\hbar \omega \hat a^+\hat a;
\end{equation}
here $\hat a=d/dz, \hat a^+ =z$ (because they act on functions $f(z)$).

The spectrum of operator $\hat H_{Cl}$ differs from that for quantized
oscillator $E_n=\hbar \omega (n+1/2)$. We would obtained correct answer
using the Hamiltonian (20). Surprisingly, Eq. (20) automatically gives
correct ordering of operators. It is important to elucidate the nature of
the "quantum energy" $\hbar \omega/2$. Functions $Z_n$ correspond to
classical periodic motion with frequencies $\omega n$ in the $1D$
$\varphi$ space (non-equilibrium states). But there is another canonical
variable $|z|$, characterized by the Gibbs distribution, which also
contributes to the total energy. So, the "zero energy" $\hbar \omega/2$
should be attributed to the universal influence of the thermal bath.

{\bf Restoration of the equilibrium state.} To model restoration of the
Gibbs distribution one may introduce, for example, friction. The equation of
motion for harmonic oscillator then reads
\begin{equation}
\ddot q+\alpha\dot q +\omega^2 q=0,
\end{equation}
where $\alpha>0$ specifies friction. For infinitesimal $\alpha$ the general
solution of (31) is
\begin{equation}
q(t)=c_1e^{-i(\omega -i\alpha/2)t} + c_2e^{i(\omega +i\alpha/2)t},
\end{equation}
Evidently, there is only one equilibrium  distribution.
We see that the classical deterministic motion disappears in the limit
$t\rightarrow\infty$ ($t>t_r\sim \alpha^{-1}$). It means that in this model
of quantum oscillator all the probability amplitudes tend to zero when
$t\rightarrow\infty$ .

\section {\bf Linear chain of harmonic oscillators in a heat bath}

It is not difficult to show that the linear chain of harmonic oscillators
in a heat bath models standard $1D$ quantum field theory. Again, it is the
Gibbs distribution that is responsible for the non-trivial measure (see (17),
(25)) in the Fock space.

Hamiltonian of a chain of harmonic oscillators can be written in the form
\begin{equation}
H=\frac{1}{2}\sum_n(\frac{p_n^2}{m}+\tilde\gamma(q_n
-q_{n-1})^2 +\gamma q_n^2) \ \ (\tilde\gamma, \gamma >0).
\end{equation}
In normal coordinates $a(k),a^*(k)$ it reads
\begin{equation}
H=\frac{1}{2}\int\limits_{-\Delta}^{\Delta}dk\omega(k)
\left(a^*(k) a(k)+ a(k) a^*(k)\right),
\end{equation}
where $$a(-k)=\frac{1}{\sqrt 2}\left(u(k)\sqrt {m\omega(k)}+i\frac{p(k)}
{\sqrt {m\omega(k)}}\right),$$
\begin{equation}
q_n=\int\limits_{-\Delta}^{\Delta}dku(k)\varphi^*_n(k),\ \
p_n=\int\limits_{-\Delta}^{\Delta}dkp(k)\varphi_n(k), \ \
\varphi_n(k)=\frac{1}{\sqrt{2\Delta}}e^{\frac{i\pi n}{\Delta}k},\ \
\Delta=\frac{\pi}{a};
\end{equation}
here $a$ is the distance between the neighbour oscillators,
$\omega^2(k)=\gamma/m +4(\tilde\gamma/m)\sin^2(\pi k/2\Delta)$. In the limit
$a\rightarrow 0$, $n\rightarrow \infty$,
$an\rightarrow x$, $a^2\tilde\gamma/m \rightarrow 1$, $\gamma/m =M^2$,
$q_n\sqrt {m/a}\rightarrow \varphi(x,t)$, one obtains the $1D$ theory of free
scalar field $\varphi$ with mass $M$. It is important that\\
1) the Hamiltonian (34) presents a set of noninteracting oscillators with
frequencies $\omega(k)$,\\
2) this theory is Lorentz invariant (the Lagrangian is $(\dot \varphi^2 -
\varphi^{\prime 2} -M^2\varphi^2)/2$), so all the fields $a(k),a^*(k)$ with
different $k$ transform one into another by the Lorentz transformation.

Evidently, the Gibbs measure $\sim \exp(-\beta\int dk\omega(k)a^*(k) a(k)$
differs from the Fock space measure $\exp(-\int dka^*(k) a(k)/\hbar)$. But
in fact the difference is formal. Introducing new variables $\tilde
a(k)=\sqrt{\lambda_k} a(k),\ \ \lambda_k=\omega(k)/\omega$,
$\omega=\omega(0)$, one rewrites  (34) \begin{equation}
H=\omega\int\limits_{-\Delta}^{\Delta}dk \tilde a^*(k)\tilde a(k).
\end{equation}
The Gibbs measure now takes the form
\begin{equation}
\prod\limits_{k} \lambda_k^{-1}\beta\omega\frac{d\tilde a^*(k)\wedge d\tilde
a(k)}{2\pi i}e^{-\beta\omega\int dk\tilde a^*(k)\tilde a(k)}
\end{equation}
(the integration limits are omitted). The infinite factor $\prod
\lambda_k^{-1}$ is not essential --- it is eliminated by normalization.
The corresponding change of functionals
$\Phi[a]\rightarrow \Phi[\lambda^{-1/2}\tilde a]$ leads to redefinition
$\Phi[\lambda^{-1/2}\tilde a]=\tilde \Phi[\tilde a]$. Thus, we come to the
scalar product for complex valued functionals $\Phi_1,\Phi_2$ in the Fock
space
\begin{equation}
(\Phi_1,\Phi_2)=\int\prod\limits_{k}\frac{d a^*(k)\wedge da(k)}{ih} e^{-\int
dka^*(k) a(k) /\hbar} \overline{\Phi_1[a]}\Phi_2[a]
\end{equation}
(in the limits $a\rightarrow 0$, $\Delta\rightarrow \infty$). From (38) it
follows that $\hat a^+(k)\Phi[a^*]=a^*(k) \Phi[a^*],\ \ \hat a(k)
\Phi[a^*]=\hbar\delta\Phi[a^*]/\delta a^*(k)$, and
$[\hat a(k),\hat a^+(k^{\prime})]=\hbar\delta(k-k^{\prime})$.

This construction elucidates concepts of space, quantum mechanics and quantum
fields. The physical space is nothing but the chain of harmonic oscillators,
fields are excitations of the chain, particles (quanta) are one-particle
excitations of the fields, and quantum theory appears as a method
appropriate for description of non-equilibrium distributions characterizing
the system. One should distinguish "the physical space" from the outer
one in which it is embedded. The physical space is the linear chain of
oscillators. Excitations of the oscillators ("matter") propagate in the
physical space.  The outer space is analogous to the multidimensional space
($10D $ or $26D$) in superstring theory. The $3D$ physical space also can be
modeled in this way as a $3D$ network made of superstrings [2].

It is important that here the neighbours of an oscillator are
responsible for taking it out from the equilibrium. The Planck constant
$\hbar=1/\beta\omega$ is universal for all the oscillators, i.e.  it is a
fundamental constant of the theory.  Distributions
\begin{equation}
\overline{\Phi[a]}\Phi[a]e^{-\int dk a^*(k)a(k)/\hbar}
\end{equation}
describe non-equilibrium states of the system (because they differ
from the equilibrium one). To describe the evolution of density
$\rho[a^*,a]=\overline{\Phi[a]}\Phi[a]$ it is enough to describe the
evolution of $\Phi[a]$, i.e. the evolution of non-equilibrium states is
described by probability amplitudes (wave functionals) $\Phi[a]$. In this
model the Planck constant $h$ and the wave functions (functionals) appear
simultaneously. The chain of harmonic oscillators models also the bosonic
string.

\section {\bf Conclusion}

It was shown that small deviations from equilibrium states of harmonic
oscillator in a thermal bath are described by probability amplitudes (in the
time interval $t_r>t\gg \omega^{-1}$). Thus, quantum mechanics can be modeled
in framework of classical theory. Of course, it does not mean that one can
obtain quantum theory of e.g. electron from classical theory of a pointlike
particle. The result is meaningful only on the fundamental level of the
Planck scales. For example, a chain of such oscillators in a thermal bath
models one-dimensional quantum field theory (or quantized bosonic string).
Excitations of the chain (quanta, particles) move along the chain
--- they model "matter". In this case, one simultaneously models both the
$1D$ physical space and quantum mechanics.

Another important feature of the model --- natural appearance of the Planck
constant $h$ (see (11), (17)). The probabilities, probability amplitudes  and
constant $h$ appear as consequences of putting oscillator into a heat
bath. The other attributes of QM (operators, complex functions) are
indispensable features of CM. Actually, there is no need to introduce a
thermal bath by hand. Complex ergodic systems are stochastic by themselves,
i.e probabilities pervade Hamiltonian mechanics, and the Gibbs distribution
also follows from CM. The "analogy of the evolution of classical correlation
functions with quantum mechanics" was discussed in [12] (without
introducing $h$; see also [13]).

It follows from (32) that in the (1+1) universe, modeled by a chain of
harmonic oscillators, quantum description will be actual only in the time
interval ($t_r,\omega^{-1}$); here $t_r$ is the relaxation time of the chain
excitations (if it doesn't exceed that of the oscillators).  When
$t\rightarrow\infty$ the system tends to the equilibrium state ("vacuum").
Any coherent motion dies away. This is another kind of "death" of the
"universe" in contrast with the "thermal death" of XIX-th century. In the
latter case matter becomes thermalized, while here matter "disappears"
because this is nothing but quantum excitation of the chain (i.e. vacuum)
specified by some probability amplitudes.


\begin{thebibliography} {50}

\bibitem [1] {1} 't Hooft G 1999 {\it Class. Quant. Grav.} {\bf 16}
3263-3279

\bibitem [2] {2} Prokhorov L V 2003 {\it Quantum mechanics --- problems and
paradoxes} (SPb: NIIKh SPbSU) (in Russian); 2004 {\it Phys. Atom. Nucl.}
{\bf 67} 1299-1311

\bibitem [3] {3} Markopoulou F, Smolin L {\it Quantum Theory from
Quantum Gravity} Preprint gr-qc/0311059

\bibitem [4] {4} Minic D, Tze C H 2003 {\it Phys. Rev.} {\bf D 68}, 061501-5;
{\it What is Quantum Theory of Gravity?} Preprint hep-th/0401028

\bibitem [5] {5} Koopman B O 1931 {\it Proc. Nat. Acad. Sc. USA} {\bf 17}
315-318

\bibitem [6] {6} Strocchi F 1966 {\it Rev. Mod. Phys.} {\bf 38} 36-40

\bibitem [7] {7} Mc Ewan J 1993 {\it Found. Phys.} {\bf 23} 313-327

\bibitem [8] {8} Landau L D, Lifshiz E M 1980 {\it Statistical Physics}
(Oxford: Pergamon Press), Part I

\bibitem [9] {9} v. Neumann J 1932 {\it Ann. Math.} {\bf 33} 587-642

\bibitem [10] {10} Bargmann V 1961 Comm. Pure. Appl. Math. {\bf 14} 187-214

\bibitem [11] {11} Hurt N E 1983 {\it Geometric Quantization in Action}
(D.~Reidel Publishing Company, London)

\bibitem [12] {12} Wetterich C 1997 {\it Phys. Lett.} {\bf B399} 123-129

\bibitem [13] {13} Morgan P 2004 {\it Phys. Lett.} {\bf A321} 216-224

\end{thebibliography}
\end{document}